\documentclass{ifacconf}
\renewcommand{\baselinestretch}{1}
\pdfoutput=1
\usepackage{natbib}        
\usepackage{amsmath,amssymb,amsfonts}
\usepackage[hyphens]{url}
\usepackage{marginnote}
\usepackage[dvipsnames]{xcolor}
\usepackage{amsmath}
\usepackage{xfrac}
\usepackage{amssymb}
\usepackage{units}
\usepackage{graphicx}
\usepackage{booktabs}
\usepackage{mathtools}

\usepackage{verbatim}
\usepackage{dsfont}
\usepackage[acronym]{glossaries}
\usepackage{lettrine}
\usepackage{lipsum}
\usepackage[ruled,vlined,algo2e]{algorithm2e}

\newif\ifmargincomments
\margincommentstrue

\ifmargincomments

\else

\fi

\newif\ifrev
\revfalse 

\ifmargincomments

\else

\fi

\ifrev
\newcommand{\rev}[1]{{\leavevmode\color{blue}#1}}
\else
\newcommand{\rev}[1]{{\leavevmode#1}}
\fi

\newcommand{\flexbrac}[1]{\if\relax\detokenize{#1}\relax \else (#1) \fi}
\newcommand{\flexcomma}[1]{\if\relax\detokenize{#1}\relax \else ,#1 \fi}

\newcommand{\cG}{\mathcal{G}}

\newcommand{\cR}{\mathcal{R}}

\newcommand{\cV}{\mathcal{V}}

\newcommand{\sR}{\mathbb{R}}

\definecolor{lightblue}{rgb}{0.60784,0.76078,0.90196}
\definecolor{darkblue}{rgb}{0.26667,0.44706,0.76863}
\definecolor{lightgreen}{rgb}{0.66275,0.81569,0.55686}
\definecolor{darkgreen}{rgb}{0.43922,0.67843,0.27843}
\definecolor{orange}{rgb}{0.92941,0.49020,0.19216}
\definecolor{yellow}{rgb}{1.00000,0.75294,0.00000}
\definecolor{grey}{rgb}{0.64706,0.64706,0.64706}
\definecolor{purple}{rgb}{0.51373,0.23529,0.04706}
\newacronym{abk:mod}{MoD}{Mobility-on-Demand}
\newacronym{abk:amod}{AMoD}{Autonomous Mobility-on-Demand}
\newacronym{abk:iamod}{\mbox{I-AMoD}}{intermodal \gls{abk:amod}}
\newacronym{abk:bpr}{BPR}{Bureau of Public Roads}
\newacronym{abk:bev}{BEV}{Battery Electric Vehicle}
\newacronym{abk:ca}{CA}{congestion-aware}
\newacronym{abk:cara}{CARS}{congestion-aware routing scheme}
\newacronym{abk:ffcs}{FFCS}{free floating car sharing systems}
\newacronym{abk:ghg}{GHG}{greenhouse gas}
\newacronym{abk:icev}{ICEV}{ Internal Combustion Engine Vehicle}
\newacronym{abk:kpi}{KPIs}{Key Performance Indicators}
\newacronym{abk:lw}{LW}{Lightweight}
\newacronym{abk:lp}{LP}{Linear Program}
\newacronym{abk:mcfp}{MCFP}{multi-commodity flow problem}
\newacronym{abk:nyc}{NYC}{New York City}
\newacronym{abk:spp}{SPP}{shortest path problem}
\newacronym{abk:kdspp}{k-dSPP}{k-disjoint \gls{abk:spp}}
\newacronym{abk:su}{SU}{Sport Utility}

\newcommand{\Jacc}{J_\mathrm{acc}}
\newcommand{\Jtopt}{J_\mathrm{time}}

\newcommand{\Pm}{P_m}

\newcommand{\am}{\alpha_m}

\newcommand{\dm}{d_m}

\newcommand{\fm}{f^m}
\newcommand{\nr}{n_r}
\newcommand{\npop}{n_\mathrm{pop}}

\newcommand{\om}{o_m}

\newcommand{\tth}{T_\mathrm{max}}
\newcommand{\tmp}{t^m_p}

\newcommand{\aur}{u_r}

\newcommand{\xr}{x^\mathrm{R}}

\newcommand{\xoptm}{x^{m\star}}
\newcommand{\xbar}{\bar{x}}
\newcommand{\xm}{x^m}
\newcommand{\xbpm}{\xbar^{m,p}}
\newcommand{\setOfArcs}{\mathcal{A}}

\newcommand{\setOfArcsCar}{\mathcal{A}_{\mathrm{C}}}
\newcommand{\setOfArcsPT}{\mathcal{A}_{\mathrm{P}}}
\newcommand{\setOfArcsPedestrian}{\mathcal{A}_{\mathrm{W}}}
\newcommand{\setOfArcsBikes}{\mathcal{A}_{\mathrm{B}}}
\newcommand{\setOfArcsSwitch}{\mathcal{A}_{\mathrm{S}}}

\newcommand{\GraphCar}{\mathcal{G}_\mathrm{C}}
\newcommand{\GraphPT}{\mathcal{G}_\mathrm{P}}
\newcommand{\GraphBikes}{\mathcal{G}_\mathrm{B}}
\newcommand{\GraphPedestrian}{\mathcal{G}_\mathrm{W}}
\newcommand{\GraphOrigins}{\cG_\mathrm{O}}
\newcommand{\GraphDestinations}{\cG_\mathrm{D}}

\newcommand{\Jquanti}{J_\mathrm{acc,paths}}

\newcommand{\Mr}{\mathcal{M}_r}

\newcommand{\ncarsmax}{N_\mathrm{cars,max}}
\newcommand{\One}{\mathds{1}}

\newcommand{\setOfVertices}{\mathcal{V}}

\newcommand{\setOfVerticesCar}{\mathcal{V}_{\mathrm{C}}}
\newcommand{\setOfVerticesPT}{\mathcal{V}_{\mathrm{P}}}
\newcommand{\setOfVerticesBikes}{\mathcal{V}_{\mathrm{B}}}
\newcommand{\setOfVerticesPedestrian}{\mathcal{V}_{\mathrm{W}}}
\newcommand{\setOfVerticesOrigins}{\mathcal{V}_\mathrm{O}}
\newcommand{\setOfVerticesDestinations}{\mathcal{V}_\mathrm{D}}
\newcommand{\regreba}{\gamma_\mathrm{R}}
\newcommand{\regtime}{\gamma_\mathrm{time}}

\newcommand{\epsm}{\varepsilon_m}

\newcommand{\arc}{(i,j)}

\begin{document}
\begin{frontmatter}
	
	\title{On Accessibility Fairness in\\ Intermodal Autonomous Mobility-on-Demand Systems\thanksref{footnoteinfo}} 
	
	\thanks[footnoteinfo]{This publication is partially supported by the project NEON (with project number 17628 of the research program Crossover, which is (partly) financed by the Dutch Research Council (NWO)), and by Eindhoven Artificial Intelligence Systems Institute (EAISI).}
	
	\author[First]{Mauro Salazar},
	\author[Second]{Sara Betancur Giraldo},
	\author[First]{Fabio Paparella},
	\author[First]{Leonardo Pedroso}
	
	\address[First]{MOVEMENT Research Group, Control Systems Technology section, Eindhoven University of Technology, 
		5600 MB, Eindhoven, The~Netherlands. \\E-mail: \{m.r.u.salazar,f.paparella,l.pedroso\}@tue.nl.}
	\address[Second]{E-mail: s.betancur.giraldo@student.tue.nl}

\begin{abstract}
	Research on the operation of mobility systems so far has mostly focused on minimizing cost-centered metrics such as average travel time, distance driven, and operational costs. Whilst capturing economic indicators, such metrics do not account for transportation justice aspects.	In this paper, we present an optimization model to plan the operation of Intermodal Autonomous Mobility-on-Demand (I-AMoD) systems, where self-driving vehicles provide on-demand mobility jointly with public transit and active modes, with the goal to minimize the accessibility unfairness experienced by the population. Specifically, we first leverage a previously developed network flow model to compute the I-AMoD system operation in a minimum-time manner. Second, we formally define accessibility unfairness, and use it to frame the minimum-accessibility-unfairness problem and cast it as a linear program. We showcase our framework for a real-world case-study in the city of Eindhoven, NL. Our results show that it is possible to reach an operation that is on average fully fair at the cost of a slight travel time increase  compared to a minimum-travel-time solution. Thereby we observe that the accessibility fairness of individual paths is, on average, worse than the average values obtained from flows, setting the stage for a discussion on the definition of accessibility fairness itself.
\end{abstract}´


\end{frontmatter}
\section{Introduction}
\begin{figure}[t!]
	\centering
	\includegraphics[trim={0 0 0 0},clip,width=0.9\linewidth]{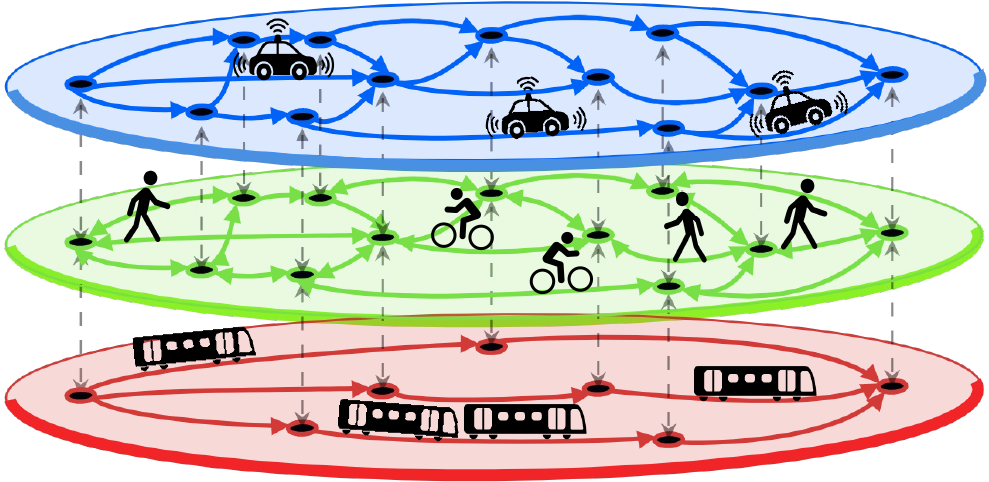}
	\caption{Schematic I-AMoD network representation. Adapted from~\cite{Wollenstein-BetechSalazarEtAl2021}.}
	\label{fig:digraph}
\end{figure}

\lettrine{M}{obility} systems are currently undergoing an important transformation with the advent of disruptive technologies such as autonomous driving, connectivity, and powertrain electrification, enabling the deployment of new mobility paradigms, such as \gls{abk:amod} systems, whereby on-demand mobility is provided by self-driving cars.
Yet, if deployed on their own and with the wrong purpose, such systems may actually backfire and not be beneficial to society~\citep{Burns2013}.
At the same time, in recent works, we showed that when combined with public transit and active modes, such \gls{abk:iamod} systems, see Fig.~\ref{fig:digraph}, could significantly improve the mobility system performance in terms of travel time and operational cost~\citep{SalazarLanzettiEtAl2019}, also in the presence of self-interested users~\citep{Wollenstein-BetechSalazarEtAl2021}.
However, these and other studies are focused on 
conventional metrics such as travel time and cost that fail in explicitly capturing the main purpose of mobility systems: to provide accessibility~\citep{Martens2017}.
Against this backdrop, in this paper we devise a network flow optimization framework to study accessibility-oriented planning of \gls{abk:iamod} systems, whereby we provide a possible quantitative definition of accessibility unfairness and minimize it.
\subsubsection{Literature Review:}
Our work pertains to two research streams: \gls{abk:amod} models and transport justice.

There exist several approaches to study
\gls{abk:amod} systems~\citep{IglesiasRossiEtAl2017,LevinKockelmanEtAl2017,ZardiniLanzettiEtAl2022}.
In particular, multi-commodity network flow models~\citep{PavoneSmithEtAl2012,SpieserTreleavenEtAl2014} are ideal for planning and design optimization purposes as they do not scale with the number of vehicles, and allow for the implementation of multiple objectives and constraints.
To date, they have been successfully applied to study congestion-aware operations~\citep{RossiZhangEtAl2017,SalazarTsaoEtAl2019,SoloveySalazarEtAl2019}, ride-pooling~\citep{TsaoMilojevicEtAl2019,PaparellaPedrosoEtAl2024b}, electric fleets~\citep{LukeSalazarEtAl2021,PaparellaChauhanEtAl2024} and their interactions with the power grid~\citep{RossiIglesiasEtAl2018b,TuranTuckerEtAl2019,EstandiaSchifferEtAl2021}, and intermodality~\citep{SalazarLanzettiEtAl2019,Wollenstein-BetechSalazarEtAl2021}.
Yet all these papers optimize the operation of the system with the minimization of time and/or operational cost as an objective, and do not account for accessibility or transport justice. 

On transport justice,~\cite{Sheller2018} investigated the relations between diversity of groups of people and uneven mobility, whilst in the context of AI and data related to governance,~\cite{ServouBehrendtEtAl2023} argued about the importance of hybrid governance, i.e., the relation between algorithms and humans to reconfigure mobility towards sustainability, avoiding reproduction of political and social inequalities on decision making processes, whilst ensuring public value.
Although these research works are insightful on issues related to transportation systems and their impacts on population welfare, these approaches lack quantification and indicators that reveal where adjustments can be made to improve accessibility of a certain population.
In this context,~\cite{HoldenLinnerudEtAl2013} proposed quantitative indicators that assess the sustainability of the transportation sector in relation to the key dimensions of sustainable development from a macroscopic perspective.
Additionally, on transportation planning based on principles of justice, \cite{Martens2017} proposes two indices to quantify accessibility: potential mobility and accessibility fairness.
Our objective is to adapt these indicators to include them within mathematical optimization frameworks for mobility systems with accessibility fairness as the main objective.
In this context, \cite{BangDaveEtAl2024} recently proposed a strategic routing game for car-only transportation systems aimed at maximizing mobility equity and presented illustrative results for a small toy example network.

In conclusion, to the best of the authors' knowledge, there are no frameworks to optimize the operation of intermodal mobility systems with transport justice objectives in a scalable fashion.

\subsubsection{Statement of Contributions:}
This paper aims at providing a first optimization problem formulation and solution algorithm to study the operation of \gls{abk:iamod} systems in a way that goes beyond conventional time-minimization paradigms, and focuses on accessibility fairness metrics with a twofold contribution:
First, we provide a quantitative definition of (average) accessibility fairness, and formulate the minimum-accessibility-unfairness operational problem as a \gls{abk:lp} that can be solved in polynomial time with global optimality guarantees.
Second, we devise a polynomial-time algorithm to reconstruct paths in a minimum-accessibility-unfairness fashion, given a set of flows. 
We showcase our framework studying the deployment and operation of an \gls{abk:iamod} system in the city of Eindhoven, NL.

\subsubsection{Organization:}
The remainder of this paper is structured as follows. In Section~\ref{sec:model}, we present a network flow model for the \gls{abk:iamod} and model the average accessibility unfairness in a quantitative, convex fashion. Thereby, we cast the optimal operational problem from a planning perspective as an~\gls{abk:lp}, and devise a path allocation algorithm computing the accessibility fairness level of the paths stemming from the optimal flows.
We showcase our framework with numerical results for Eindhoven, NL, in Section~\ref{sec:results}, whilst drawing the conclusions and providing an outlook in Section~\ref{sec:conclusion}.
\section{I-AMoD System Model and Problem Formulation}\label{sec:model}
This section presents a network flow model of the \gls{abk:iamod} system under consideration, formulates the time-optimal planning problem, followed by a convex model quantitatively capturing average accessibility unfairness, and an algorithm to compute the minimum-accessibility-unfairness paths from given flows.

\subsection{Intermodal Network Structure}

Similar to~\cite{SalazarLanzettiEtAl2019,Wollenstein-BetechSalazarEtAl2021}, we model the mobility system and its different transportation modes on a digraph $\cG=(\setOfVertices,\setOfArcs)$ consisting of a set of nodes $\setOfVertices$ and a set of arcs $\setOfArcs\subseteq\setOfVertices\times\setOfVertices$, which contain a car network layer $\GraphCar = (\setOfVerticesCar,\setOfArcsCar)$, a public transportation layer $\GraphPT=(\setOfVerticesPT,\setOfArcsPT)$,
a bicycle layer $\GraphBikes=(\setOfVerticesBikes,\setOfArcsBikes)$,
a walking layer \mbox{$\GraphPedestrian=(\setOfVerticesPedestrian,\setOfArcsPedestrian)$}, an origin layer $\GraphOrigins = (\setOfVerticesOrigins,\emptyset)$ and a destination layer $\GraphDestinations=(\setOfVerticesDestinations,\emptyset)$, which contain all origin and destination nodes, respectively, with no internal arcs.
The car network consists of intersections $i\in\setOfVerticesCar$ and car-driven road links $\arc\in\setOfArcsCar$.
We model public transportation, i.e., tram and subway lines as distinct trees, using a set of station nodes $i\in\setOfVerticesPT$ and a set of line segments $\arc\in\setOfArcsPT$.
The bicycles and walking network consist of bike paths and walkable streets $\arc\in\setOfArcsBikes$ and $\arc\in\setOfArcsPedestrian$ between their intersections $i\in\setOfVerticesBikes$ and $i\in\setOfVerticesPedestrian$, respectively.
Finally, we model the possibility of customers switching transportation modes (e.g., exiting \rev{public transport} or hailing an \gls{abk:amod} ride) by connecting the origin layer to \rev{walking}, car, and bicycles layer, walking \rev{to} bike, public transportation and car layer, and car, walking and biking layer to the destination layers.
We capture all interconnections with a set of mode-switching arcs  $\setOfArcsSwitch\subseteq\setOfVerticesCar\times\cV_\mathrm{D}\cup \cV_\mathrm{O}\times\setOfVerticesCar \cup \setOfVerticesPT\times\setOfVerticesPedestrian \cup\setOfVerticesPedestrian \times\setOfVerticesPT\cup\setOfVerticesBikes\times\setOfVerticesPedestrian \cup \cV_\mathrm{O} \times \setOfVerticesPedestrian \cup \setOfVerticesPedestrian \times \cV_\mathrm{D} \cup \cV_\mathrm{O} \times \setOfVerticesBikes \cup \setOfVerticesBikes \times \cV_\mathrm{D} $. Additionally, layers do not share nodes with each other. 
Accordingly, $\setOfVertices = \setOfVerticesPedestrian\cup\setOfVerticesBikes\cup \setOfVerticesCar \cup \setOfVerticesPT \cup \cV_\mathrm{O} \cup \cV_\mathrm{D}$ and $\setOfArcs = \setOfArcsPedestrian \cup\setOfArcsBikes\cup \setOfArcsCar \cup \setOfArcsPT \cup \setOfArcsSwitch$ hold.

We consider a set of $R$ regions $r\in\cR=\{1,\dots,R\}$ with population $\nr$ so that $\sum_{r\in\cR} \nr = \npop$. Considering a time of day (e.g., morning rush hour), each region has a set of mobility demands to and/or from relevant locations (e.g., from a household to a workplace, school, hospital, supermarket, etc.), formally defined by the tuple $(\om,\dm,\am)$ with $m\in\Mr$, containing origin and destination nodes $\om,\in\cV_\mathrm{O},\dm\in\cV_\mathrm{D}$ and number of users per unit time $\am$.

Assuming no endogenous congestion effects, we consider a constant travel time vector $t_a\geq0$ representing the time needed to traverse arc $a\in\setOfArcs$, which \rev{captures} the time to move (e.g., by car, bike, public transit, or walking) or to change mode of transportation (e.g., parking a bike, exiting an \gls{abk:amod} ride, waiting, getting on a public transit vehicle, etc.).

\subsection{Operational Constraints and Time-optimal Problem}\label{subsec:op}
Leveraging the network defined so far, we define arc-dependent flow variables for each origin-destination pair (o-d-pair) as $x_a^m$, representing how many people related to region $r\in\cR$ and demand $m\in\Mr$ are traveling on arc $a\in\setOfArcs$ per unit time; we define \textit{rebalancing} flows of empty vehicles realigning themselves with the car-demand distribution as $\xr_a$ for $a\in\setOfArcsCar$.
Then the flow balancing constraints state that every user flow entering each node $j$ must exit it, and are formally defined as
\begin{equation}\label{eq:flowbal}
	\begin{aligned}
		\am\One_{j=\om} + \sum_{i:(i,j)\in\setOfArcs}\xm_{(i,j)} = &\am\One_{j=\dm} + \sum_{k:(j,k)\in\setOfArcs}\xm_{(j,k)}\\
		&\forall j \in\setOfVertices,\forall m\in\Mr,\forall r\in\cR,
	\end{aligned}
\end{equation}
where $\One$ is a Boolean indicator function defined as $\One_b=1$ if $b$ is true and 0 otherwise.
Similarly, the car balancing constraint is
\begin{equation}\label{eq:carbal}
	\begin{aligned}
		&\sum_{i:(i,j)\in\setOfArcsCar}\left(\xr_{(i,j)}+\sum_{r\in\cR}\sum_{m\in\Mr}\xm_{(i,j)}\right)\\
		&= \sum_{k:(j,k)\in\setOfArcsCar}\left(\xr_{(j,k)}+\sum_{r\in\cR}\sum_{m\in\Mr}\xm_{(j,k)}\right)\;\forall j \in\setOfVerticesCar.
	\end{aligned}
\end{equation}
Finally, flows are non-negative
\begin{equation}\label{eq:nonneg}
	\begin{aligned}
		\xm_{a}&\geq0\; \forall m\in\Mr, \forall r\in\cR, \forall a\in\setOfArcs\\
		\xr_{a}&\geq 0\;\forall a\in\setOfArcsCar,
	\end{aligned}
\end{equation}
and we limit the number of cars circulating to $\ncarsmax$:
\begin{equation}\label{eq:maxcars}
	\sum_{a\in\setOfArcsCar}t_a\cdot\left(\xr_{a}+\sum_{r\in\cR}\sum_{m\in\Mr}\xm_{a}\right)\leq \ncarsmax.
\end{equation}

From this, the standard time-optimal \gls{abk:iamod} network flow problem can be defined via the objective function
\begin{equation}\label{cost:topt}
	\Jtopt = \sum_{a\in\setOfArcs} t_a\cdot \sum_{r\in\cR} \sum_{m\in\Mr} \xm_a + \regreba\cdot  \sum_{a\in\setOfArcsCar} t_a \cdot \xr_a,
\end{equation}
where $\regreba$ is a small regularization term for rebalancing, as follows:

\begin{prob}[Minimum-time Operation]\label{prb:topt}
	\hphantom{the optimal flows}\\
	The optimal flows minimizing the average travel time of the population result from
	\begin{align*}
		\min_{\{\xm\}_m\in\sR^{|\setOfArcs|},\xr\in\sR^{|\setOfArcsCar|}} \;& \Jtopt\\
		\text{s.t. } &\eqref{eq:flowbal}-\eqref{cost:topt}.
	\end{align*}
\end{prob}
Problem~\ref{prb:topt} has been studied in several variations and can be efficiently solved as an~\gls{abk:lp}. Yet it myopically focuses on minimizing the average travel time of all the users, and does not capture transportation justice effects such as accessibility fairness.

In what follows, we leverage the network-and-demand structure together with the operational constraints to frame the  maximum-accessibility-fairness problem for the \gls{abk:iamod} system under consideration.

\subsection{Accessibility Fairness Definition}\label{subsec:af}
Whilst the time-optimal operation problem has been extensively studied in the past, it is not aligned with the main purpose of mobility systems: to provide accessibility.
In line with~\cite[Chpt.\ 8]{Martens2017}, accessibility is a multi-faceted concept related to accommodating travel demands in a reasonable way.
In this paper, we explore its quantitative definition via a reasonable travel time threshold $\tth$ (e.g., \unit[20]{min}): Formally, we quantify the unaccessibility of an o-d-pair by the amount of extra time above the threshold needed to reach it via the slack variable
\begin{equation*}
	\epsm = \max\left\{0,\frac{t^\top \xm}{\am} - \tth\right\} \;\forall m\in\Mr,\forall r\in\cR,
\end{equation*}
which can be losslessly relaxed to a convex \rev{constraint} as 
\begin{equation}\label{eq:tacc}
	\epsm \geq  \max\left\{0,\frac{t^\top \xm}{\am}  - \tth\right\} \;\forall m\in\Mr,\forall r\in\cR,
\end{equation}
since it will be subject to minimization~\citep{BorsboomFahdzyanaEtAl2021}.
We observe that such a definition captures the accessibility level of an o-d-pair \textit{on average}. Specifically, if a fully accessible flow vector $\xm$ consists of more than one path, it could happen that some of those paths have a travel time above the threshold, whilst other paths have one below it, so that the weighted average travel time computed in~\eqref{eq:tacc} is below the threshold.
Second, we define the accessibility unfairness level $\aur$ of a region $r$ \rev{as the average of extra travel time experienced for each o-d-pair belonging to it, weighted with the frequency of occurrence:}
\begin{equation}\label{eq:aur}
	\aur = \frac{\sum_{m\in\Mr}\am\cdot\epsm}{\sum_{m\in\Mr}\am}\;\forall r\in\cR.
\end{equation}


Finally, we capture the total population-weighted accessibility unfairness with the cost function
\begin{equation}\label{cost:acc}
	\Jacc = \frac{\sum_{r\in\cR}\nr\cdot\aur}{\sum_{r\in\cR}\nr},
\end{equation}
and frame the maximum-accessibility problem as follows:

\begin{prob}[Minimum-accessibility-unfairness I-AMoD]\label{prb:acc}
	\hphantom{the}\\
	The flows minimizing accessibility unfairness result from
	\begin{align*}
		\min_{\{\xm\}_m\in\sR^{|\setOfArcs|},\xr\in\sR^{|\setOfArcsCar|}} \;& \Jacc + \regtime\cdot\Jtopt \\
		\text{s.t. } &\eqref{eq:flowbal}-\eqref{cost:acc},
	\end{align*}
where $\regtime$ is a small regularization term to avoid \mbox{self-loops} that can happen for accessible o-d-pairs.
\end{prob}
Problem~\ref{prb:acc} is an~\gls{abk:lp} that can be efficiently solved with global optimality guarantees with off-the-shelf solution algorithms.

\subsection{Discussion}
A few comments are in order.
First, we assume a constant travel time and no endogenous congestion effects. Such an assumption is in order for non car-based modes, and for relatively small fleets, \rev{whereby it readily captures} exogenous congestion effects~\citep{Rossi2018}. Whilst the focus of the present paper is not on car-centric mobility, endogenous-congestion-aware traffic models could be included in future extensions~\citep{SalazarTsaoEtAl2019,Wollenstein-BetechSalazarEtAl2021}.
Second, we quantitatively capture accessibility unfairness in~\eqref{eq:aur} using the flow-weighted extra time experienced by each region to accomplish its relevant trips.
As accessibility is in general a \textit{qualitative} multi-faceted concept from social science~\citep{Martens2017}, we plan to study and possibly extend its quantitative definition to capture other aspects in future works.
Finally, the slack variable definition in~\eqref{eq:tacc} captures the unaccessibility level of an o-d-pair on average, which corresponds to the path-based accessibility if there is only one path per flow-variable, which may not always be the case due to the presence of constraint~\eqref{eq:maxcars}.
In this paper, we optimize the accessibility of paths \textit{a posteriori} with the maximum-accessibility path allocation algorithm detailed in Section~\ref{subsec:pa} below.

\subsection{Path Allocation}\label{subsec:pa}
In this section, we formulate the problem of allocating optimal flows $\xoptm$ obtained as described in Section~\ref{subsec:op} above to paths by minimizing the unaccessibility level of each origin-destination pair $m$.
Given an origin node $\om$, we define destination node $\dm$ as accessible if the travel time of the path connecting them is below a maximum threshold $\tth$.
For each o-d-pair, we first construct a smaller digraph only consisting of the arcs traveled by the optimal flows as $\cG_m = (\setOfVertices_m,\setOfArcs_m)$ where $a=(i,j)\in\setOfArcs_m \iff \xoptm_{a}>0$ and $j\in\setOfVertices_m\iff \xoptm_{a}>0$ for $a=(i,j)$ or $a=(j,k)$.
Thereafter, we compute the set of all acyclic paths connecting $\om$ to $\dm$ as $\xbpm\in\{0,1\}^{|\setOfArcs_m|}$ with $p\in\{1,\dots,\Pm\}$ and $\Pm$ the total number of paths.
For each of the paths $p\in\{1,\dots,\Pm\}$, we compute the travel time from $\om$ to $\dm$ as $\tmp$, and assess their induced accessibility by verifying whether $\tmp\leq \tth$.
The path allocation problem now consists of finding the flows $\fm\in\sR^{\Pm}$ for which it holds that
\begin{equation}\label{eq:posflowstopaths}
	\begin{aligned}
		&\sum_{p=1}^{\Pm}\fm_p\cdot\xbpm=\xoptm\\
	&\fm_p\in [0,1]\;\forall p\in\{1,\dots,\Pm\},
	\end{aligned}
\end{equation}
so that accessibility is maximized.
To this end, we define the following cost-function, 
aimed at minimizing the amount of flows that are inaccessible, weighted by the quantity of time they are violating the accessibility constraint:
\begin{equation}\label{cost:quanti}
	\Jquanti^m = \sum_{p=1}^{\Pm}\max\{0,\tmp-\tth\}\cdot \fm_p,
\end{equation}
where we observe that constraints \eqref{eq:posflowstopaths} guarantee that $\sum_{p=1}^{\Pm}\tmp\cdot \fm_p = t^\top\xoptm$ for any admissible $\fm$, i.e., the average travel time of the population is always the same.

We now formulate the maximum-accessibility flow-path allocation problem as follows:
\begin{prob}[Maximum Accessibility Allocation]\label{prb:allocation}
	\hphantom{maxim}\\
	The maximum-accessibility path-flows $f_p^\star$ result from
	\begin{align*}
		\min_{\fm\in\sR^{\Pm}} \;& \Jquanti^m\\
		\text{s.t. } &\eqref{eq:posflowstopaths}, \eqref{cost:quanti}.
	\end{align*}
\end{prob}
Problem~\ref{prb:allocation} is an~\gls{abk:lp} that can be efficiently solved with off-the-shelf solution algorithms.
The full procedure is described in Algorithm~\ref{alg:paths}, where we solve Problem~\ref{prb:allocation} for each o-d-pair $m$ to compute the total accessibility unfairness level $\Jquanti^\star$ for the given optimal flows.
\begin{algorithm}[t]
	\caption{Maximum-accessibility Path Allocation}\label{alg:paths}
	\KwData{$x^\star$}
	$\Jquanti^\star\gets 0$\\
	\For{$r\in\cR$}{
	\For{$m\in\Mr$}{
		$\;\,$Compute $\cG_m = (\setOfVertices_m,\setOfArcs_m)$\\
		Compute $\{\xbpm\}_p$\\
		Solve Problem~\ref{prb:allocation}\\
		$\Jquanti^\star\gets \Jquanti^\star + \Jquanti^{m,\star}$
}
}
\KwResult{$\Jquanti^\star,f^\star$}
\end{algorithm}


\section{Results}\label{sec:results}
In this section, we study the performance achievable by an	\gls{abk:iamod} system when operated in an accessibility-aware fashion and compare it to a more standard time-optimal operation.
To this end, we compare the results obtained by solving Problem~\ref{prb:acc} and Problem~\ref{prb:topt}, i.e., minimum-accessibility-unfairness vs minimum-time operation, respectively, where we allocate optimal flows to paths via Problem~\ref{prb:allocation}, using Algorithm~\ref{alg:paths}.
We set our case-study in the city of Eindhoven, NL.
For the road network we gather information from OpenStreetMaps~\citep{HaklayWeber2008}, whilst for the public transportion network we use~\cite{GTFS2019}.
We generate realistic demand data with ALBATROSS (A Learning BAsed TRansportation Oriented Simulation System), developed in~\cite{ArentzeTimmermans2004,RasouliKimEtAl2018},
which creates o-d-demands at PC4 level, i.e., the first 4 digits of the Dutch postal code.
Specifically, it provides potential destinations and the ratio of the population that would travel there.
Multiplying such ratios with the population size \rev{$\nr$} of each PC4 region $r$ and the average number of trips per person during a day~\citep{CBS2024}, we obtain the average number of daily trips from an origin PC4 to a destination as flows $\am$ for $m\in\Mr$ and each region $r\in\cR$.
Last, we relate every o-d-pair at PC4 level to node level by evenly spreading the o-d-pairs to every node inside the origin and destination PC4.
After mild clustering (see Appendix in~\cite{PaparellaChauhanEtAl2024}), the resulting intermodal network has about 1300 nodes and 5000 arcs.
The parameters used in the case-study are described in Table~\ref{tab:one}, where we select a fleet of $\ncarsmax=\unit[4000]{cars}$, and an accessibility threshold of $\tth=\unit[20]{min}$.

Problem~\ref{prb:topt}, \ref{prb:acc} and~\ref{prb:allocation} were parsed with YALMIP~\citep{Loefberg2004} and solved with Gurobi~\citep{GurobiOptimization2021}. On a standard laptop with \unit[16]{GB} RAM and an Apple M1 Pro chip with a 10-core CPU, the first two Problems took about \unit[10]{min} to solve, whilst running Algorithm~\ref{alg:paths} took about \unit[1]{min} when parallelized over the cores.
\begin{table}[t]
	\caption{Parameters for the case-study}
	\label{tab:one}
	\begin{center}
		\begin{tabular}{lcl}
			\toprule
		  Parameter & Value &  Unit \\
		  \midrule
				$\ncarsmax$ & $4000$ & vehicles\\
			$t_a$, $a \in \cV_\mathrm{O} \times \cV_\mathrm{C}$ & $3$ & min\\
	$t_a$, $a \in \cV_\mathrm{W} \times \cV_\mathrm{P}$ & $8$ &  min\\
		$\sum_m \alpha_m$ & $59000$ &  users/hour\\
		$	\tth$ & $20$ & min\\
		\bottomrule
		\end{tabular}
	\end{center}
\end{table} 

\subsection{Performance of Minimum-accessibility-unfairness vs Minimum-travel-time I-AMoD Operation}
\begin{figure}[t!]
	\centering
	\includegraphics[trim={0 0 0 0},clip,width=\linewidth]{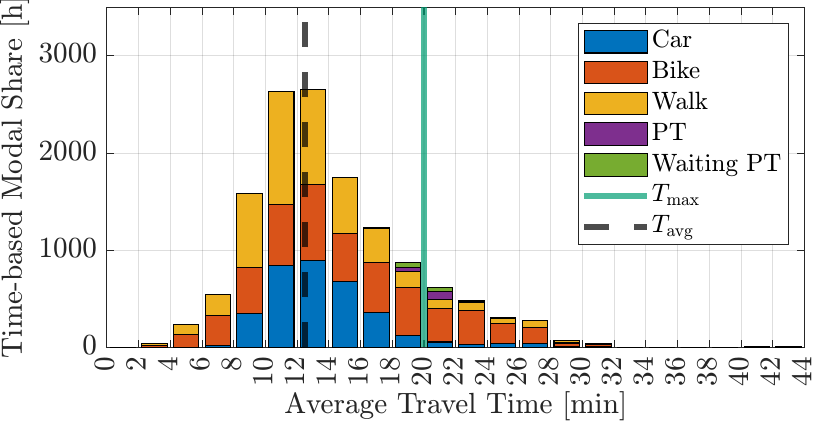}
	\includegraphics[trim={0 0 0 0},clip,width=\linewidth]{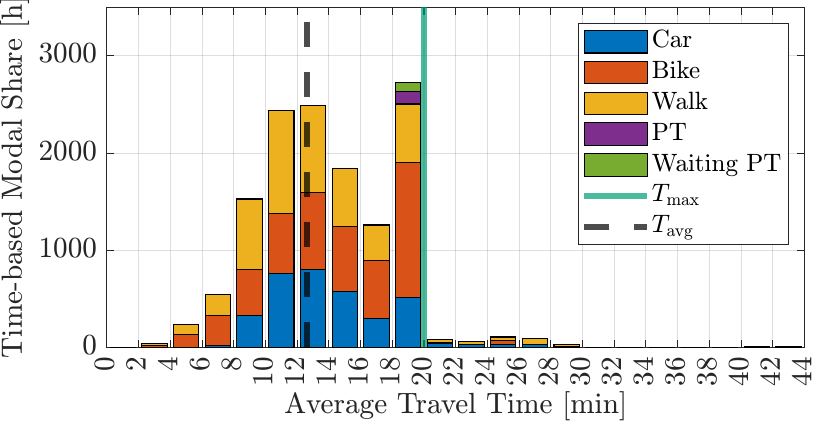}
	\caption{Comparison of time-based modal share as a function of the average travel time of user-weighted o-d-pairs for the objectives of minimum travel time (top) and minimum accessibility unfairness (bottom). \hphantom{adding more text here.}
	}
	\label{fig:odbased}
\end{figure}

\begin{figure}[t!]
\centering
\includegraphics[trim={0 0 0 0},clip,width=\linewidth]{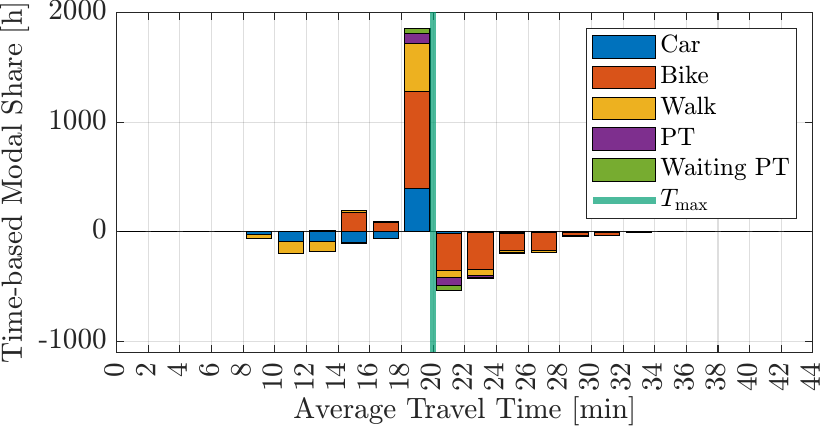}
\caption{Time-based modal share difference as a function of the average travel time of user-weighted o-d-pairs.
}
\label{fig:odbaseddiff}
\end{figure}

\begin{figure*}[t!]
	\centering
	\includegraphics[trim={0 0 0 0},clip,width=0.49\linewidth]{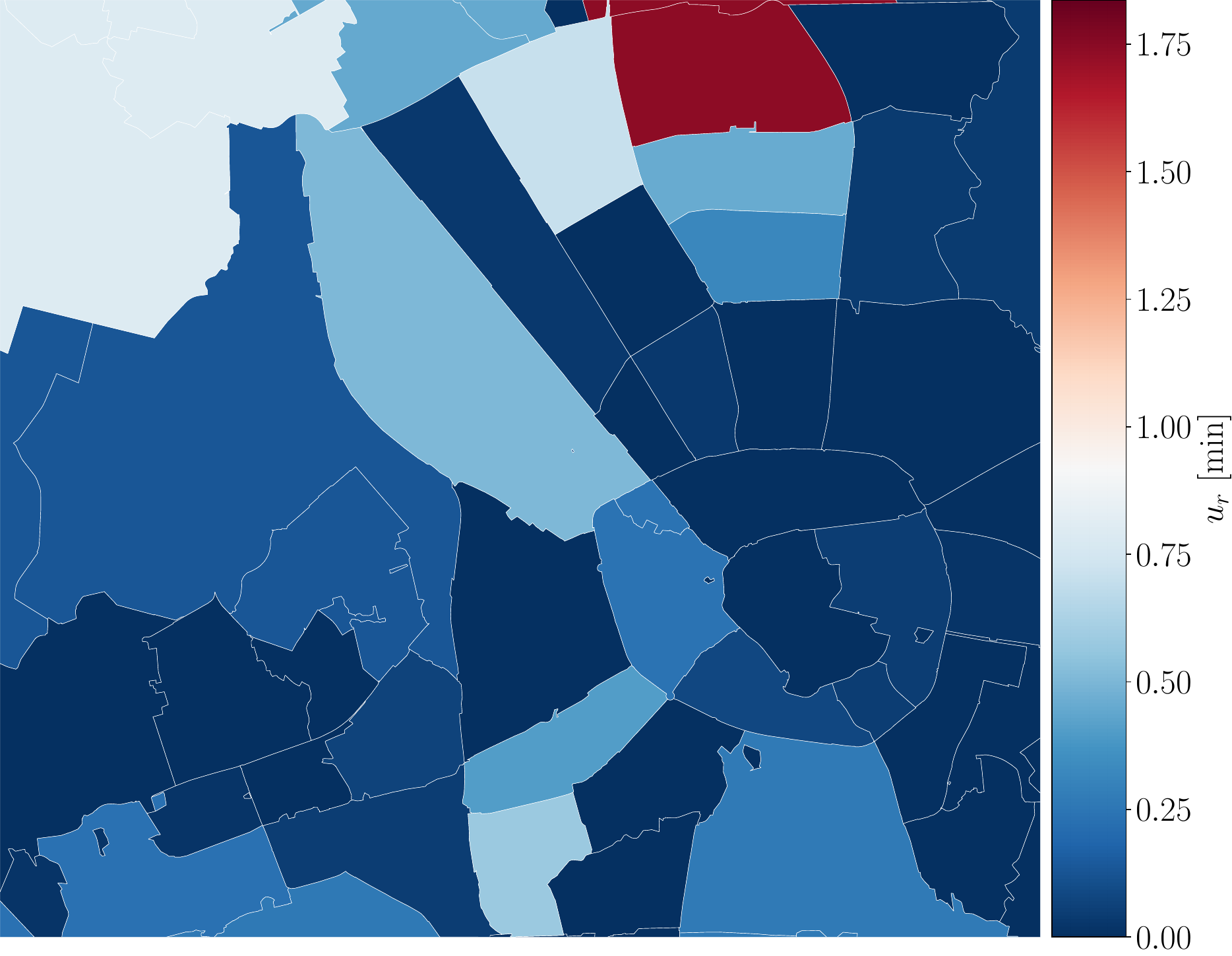}\hfill
	\includegraphics[trim={0 0 0 0},clip,width=0.49\linewidth]{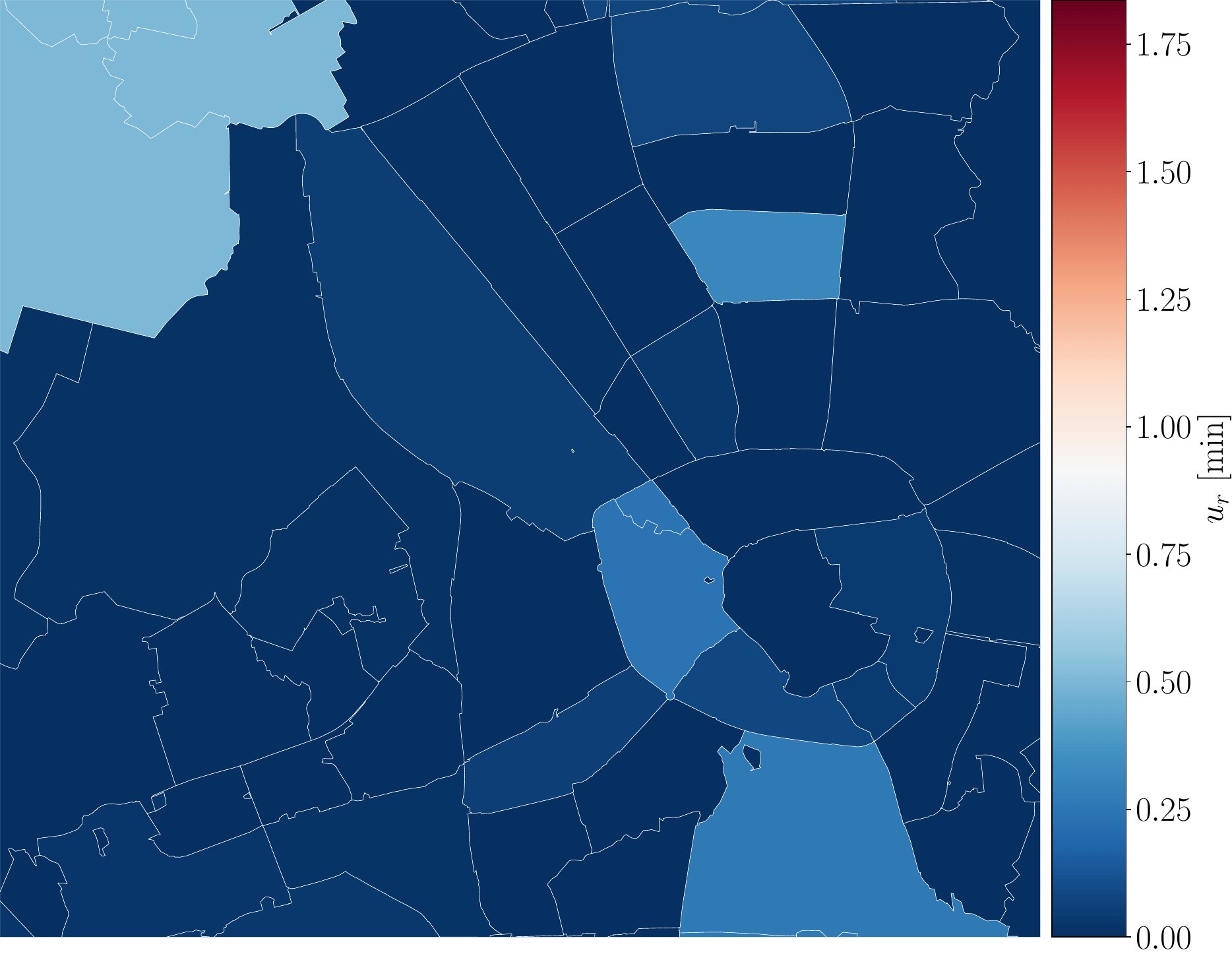}
	\caption{Average accessibility unfairness levels in the PC4 regions $u_r$ of Eindhoven, NL, for the minimum-travel-time~(left) and minimum-accessibility-unfairness operation (right).
	}
	\label{fig:heatmaps}
\end{figure*}

\begin{figure*}[t!]
	\includegraphics[trim={0 0 0 0},clip,width=0.453\linewidth]{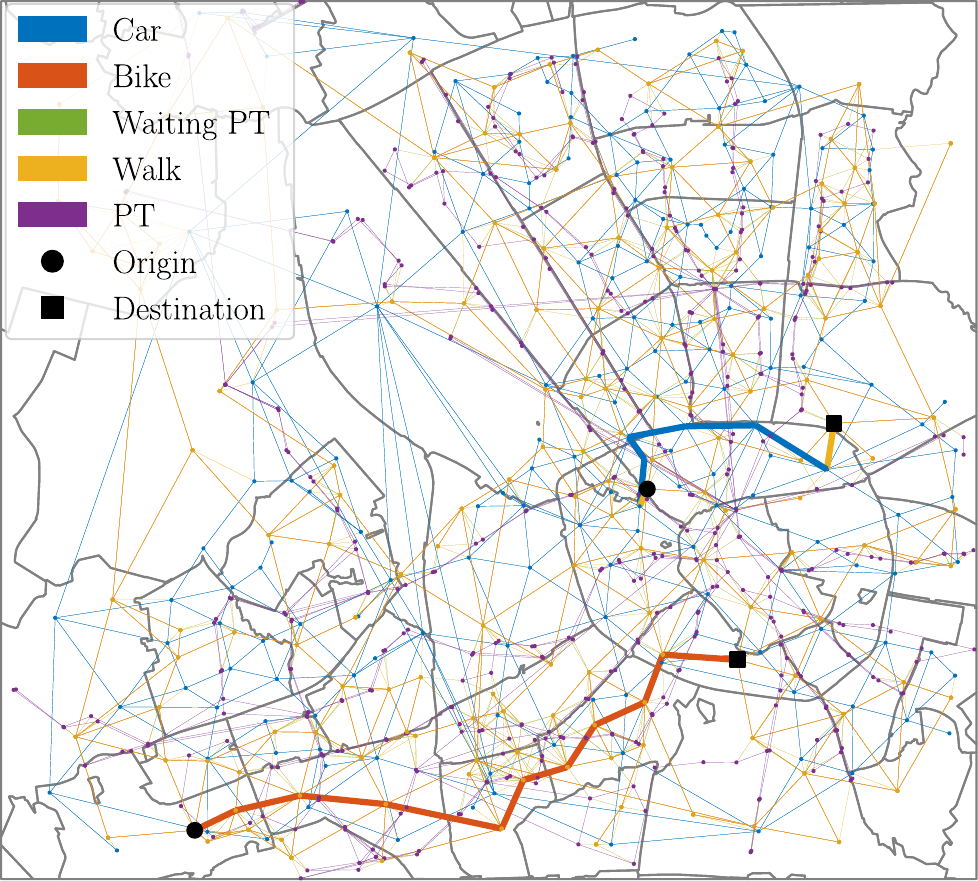}\hspace{0.9cm}
	\includegraphics[trim={0 0 0 0},clip,width=0.453\linewidth]{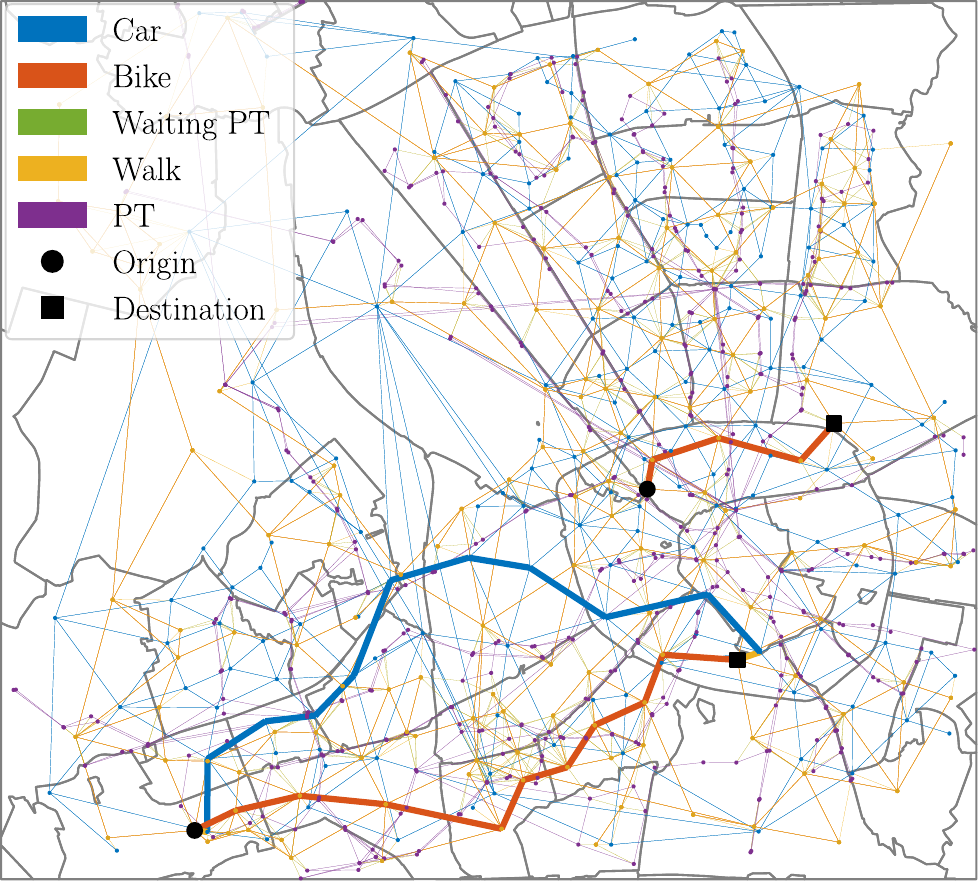}
	\caption{Minimum-travel-time (left) and minimum-accessibility-unfairness paths (right) for two o-d-pairs in Eindhoven, NL, showing that the latter operation uses cars to serve the left o-d-pair with a larger travel distance at the cost of a potentially higher rebalancing activity, whilst devoting bikes to the closer right o-d-pair within the city Ring. Thereby, the PC4 regions $u_r$ are delimited in light grey with underlying digraphs representing the different clustered modal networks of the city.
	}
	\label{fig:paths}
\end{figure*}

In this section,we investigate the differences in modal share, average travel time, and unfairness of the \gls{abk:iamod} operation obtained by solving Problem~\ref{prb:topt} and Problem~\ref{prb:acc}.
Fig.~\ref{fig:odbased} depicts the distribution of the time-based modal share over the average regional travel time distribution, whilst Table~\ref{tab:results} provides the performance achieved by the two types of operation.
\begin{table}[t]
	\vspace{1em}
	\caption{Optimization results for a fleet consisting of \mbox{$\ncarsmax=\unit[4000]{cars}$} and a time threshold of \mbox{$\tth=\unit[20]{min}$}.}
	\label{tab:results}
	\begin{center}
		\begin{tabular}{lcc}
			\toprule
			Objective & Min Time &  Max Acc \\
			\midrule
			Travel Time [min]& $12.51$ & $12.62$ \\
			Unfairness Level per o-d-pair [min]& $0.2274$ & $0.0645$ \\
			Unfairness Level per path [min]& $0.2506$ & $0.1947$ \\
			\bottomrule
		\end{tabular}
	\end{center}
\end{table}
Whilst the time-optimal operation achieves the lowest travel time, the minimum-accessibility-unfairness operation manages to shift the average travel time of each o-d-pair below the threshold $\tth$, almost completely reaching accessibility fairness at the cost of a marginal increase in average travel time below 1\%. The same result is shown in Fig.~\ref{fig:heatmaps}, depicting the average accessibility unfairness for the population of the PC4 regions $u_r$.
Fig.~\ref{fig:odbaseddiff} shows the difference between the two histograms, revealing that cars are subtracted at the cost of potentially longer rebalancing distances from already fast routes that would be accessible also by bike and other modes, and used to reduce accessibility unfairness of o-d-pairs that are further away. Fig.~\ref{fig:paths} shows this phenomenon for two exemplary o-d-pairs.
\begin{figure}[t!]
	\centering
	\includegraphics[trim={0 0 0 0},clip,width=\linewidth]{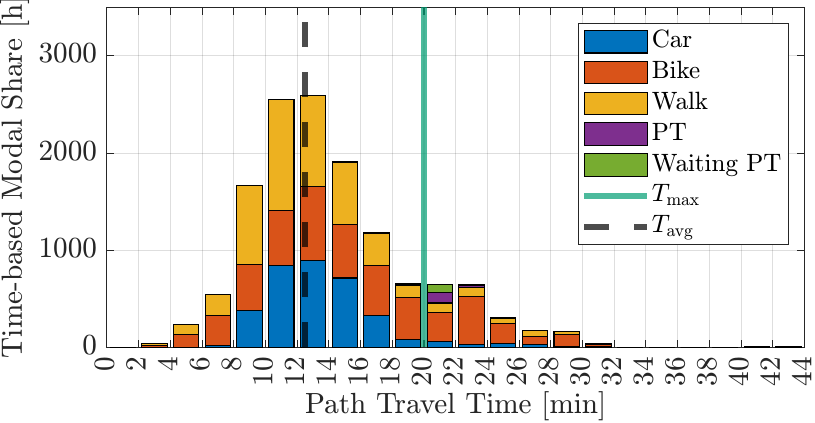}
	\includegraphics[trim={0 0 0 0},clip,width=\linewidth]{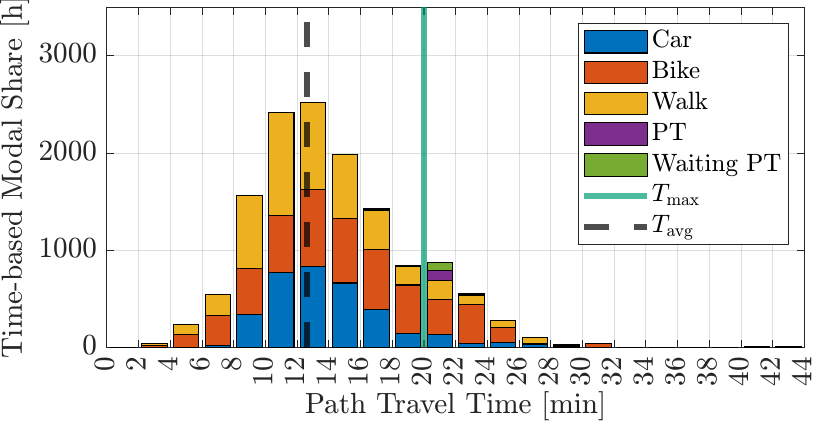}
	\caption{Comparison of time-based modal share as a function of the average travel time of user-weighted paths computed with Algorithm~\ref{alg:paths} for the different objectives of minimum travel time (top) and minimum-unfairness-accessibility (bottom).
	}
	\label{fig:pathbased}
\end{figure}
\begin{figure}[t!]
\centering
\includegraphics[trim={0 0 0 0},clip,width=\linewidth]{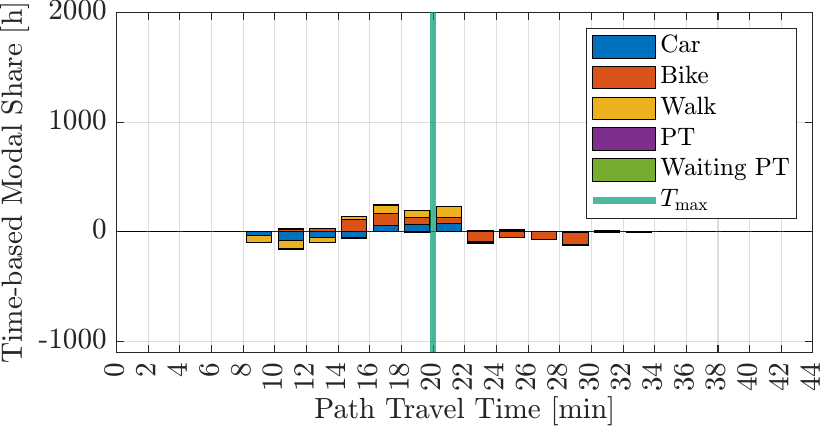}
\caption{Time-based modal share difference as a function of the average travel time of user-weighted paths.
}
\label{fig:pathbaseddiff}
\end{figure}


Interestingly, we observe a difference between the average accessibility unfairness levels achieved per o-d-pair plotted in Fig.~\ref{fig:odbased}, achieved with~\eqref{eq:tacc}--\eqref{eq:aur}, and the one computed per path with Algorithm~\ref{alg:paths}, which in general can be higher, as shown in Fig.~\ref{fig:pathbased} and~\ref{fig:pathbaseddiff}.
The reason behind this difference is that the flows $\xm$ related to o-d-pair $m$ may consist of different paths $p$ with a different travel time $\tmp$ which weighted average may be below the threshold.
This does not imply that all path-related travel times should also be below the threshold, i.e., some paths may not provide accessibility, even though their aggregate values do so on average.
Such a difference can become even larger if we further decrease $\tth$ or increase the number of cars $\ncarsmax$.
Overall, this result opens the discussion on the definition of accessibility fairness itself.
When considering travel-time-based metrics, individual trips (here, the individual paths) may have different accessibility levels  compared to their combination within repeated commuted settings (here, the o-d-pair flows). In the latter case, the lowest possible accessibility unfairness could be achieved via turn-taking mechanisms on the different paths.
\section{Conclusion}\label{sec:conclusion}
With this paper, we aimed at infusing transportation justice narratives from social science within quantitative optimization frameworks for mobility.
In particular, we devised quantitative metrics capturing accessibility fairness aspects and used them to study minimum-accessibility-unfairness operational strategies for Intermodal Autonomous Mobility-on-Demand systems and compared their performance to conventional minimum-travel-time strategies.
Our results revealed that accessibility can be maximized at the cost of a negligible average travel time increase, whilst also prompting a discussion on the definition of accessibility fairness within dynamic vs static settings.

This paper is a first step towards a more transdisciplinary perspective on mobility systems within the control and optimization community.
In the future, we would first like to capture more facets of accessibility fairness~\citep{Martens2017} with quantitative optimization models.
Second, we are interested in capturing the benefits stemming from the use of different mobility modes~\citep{TeBroemmelstroetNikolaevaEtAl2021}.
Finally, the difference observed between average and path-based accessibility prompt the study of turn-taking mechanisms to achieve system-optimal flows in an equitable~\citep{SalazarPaccagnanEtAl2021,PedrosoHeemelsEtAl2023} and/or egalitarian fashion~\citep{PedrosoAgazziEtAl2024}.

\begin{ack}
	We thank Dr. Ilse New for proofreading this paper. Furthermore, we are grateful to Prof.\ Frauke Behrendt and Prof.\ Karel Martens for the fruitful and inspiring discussions.
\end{ack}

\bibliography{../../../Bibliography/main,../../../Bibliography/mobility,../../../Bibliography/SML_papers}

\end{document}